\documentclass[10pt,letterpaper]{article}

\usepackage{opex3}
\usepackage{siunitx}

\usepackage{xcolor}
\usepackage{cite}

\begin{document}

\title{Precise shaping of laser light by an acousto-optic deflector}

\author{Dimitris Trypogeorgos,$^{1,*}$ Tiffany Harte,$^1$, Alexis Bonnin$^2$ and Christopher Foot$^1$}

\address{$^1$Clarendon Laboratory, University of Oxford, Parks Road, Oxford, OX1 3PU, UK \\ $^2$ONERA, DMPH, BP 80100, 91123, Palaiseau, France}

\email{$^*$d.trypogeorgos@physics.ox.ac.uk}

\begin{abstract}

We present a laser beam shaping method using acousto-optic deflection of light and discuss its application to dipole trapping of ultracold atoms. By driving the acousto-optic deflector with multiple frequencies, we generate an array of overlapping diffraction-limited beams that combine to form an arbitrary-shaped smooth and continuous trapping potential. Confinement of atoms in a flat-bottomed potential formed by a laser beam with uniform intensity over its central region confers numerous advantages over the harmonic confinement intrinsic to Gaussian beam dipole traps and many other trapping schemes. We demonstrate the versatility of this beam shaping method by generating potentials with large flat-topped regions as well as intensity patterns that compensate for residual external potentials to create a uniform background to which the trapping potential of experimental interest can be added. \end{abstract}

\ocis{ (020.0020) Atomic and molecular physics; (020.1475) Bose-Einstein condensates; (020.7010) Laser trapping; (230.1040) Acousto-optical devices.} 

\bibliographystyle{osajnl}
\bibliography{AODJuly2013}

\section{Introduction}

Precise shaping of laser beams is crucial to their application as confining potentials for ultracold atoms. Notable among the applications for such trapped cold atomic systems is the quantum simulation of many-particle condensed matter systems in periodic potentials~\cite{lewenstein_ultracold_2007}. Beam shaping is increasingly important too in the creation, as well as subsequent manipulation, of a Bose-Einstein condensate (BEC). Following the initial demonstration of crossing the BEC transition by purely optical means~\cite{BarrettAllOptical}, increasingly precise schemes have been developed by which to optically cool and compress an atomic cloud, including a proposed scheme whereby dynamic beam shaping techniques transform the potential between a sequence of power-laws~\cite{bruce_holographic_2011}. Minimising perturbative effects in any such experiment requires the potential experienced by the atoms to smoothly and accurately conform to a target intensity. However, accurate beam shaping becomes more challenging the greater the deviation from the diffraction-limited Laguerre-Gaussian or Hermite-Gaussian propagation modes. Additional considerations include restricting the formation of interference fringes, caused either by rapid phase variations across the beam profile or high beam coherence in an imperfect optical system. Furthermore, underlying potentials associated with the surrounding experiment may affect the confinement experienced by trapped atoms.

A prominent starting point for many quantum simulation experiments is the experimental realisation of the Bose-Hubbard Hamiltonian achieved by loading quantum degenerate bosonic atoms into an optical lattice~\cite{jaksch_cold_1998}:

\begin{equation}
	\label{eqn:BHM}
	\mathcal{H}=-J \sum_{\langle i,j \rangle} a_i^\dagger a_j+\frac{U}{2} \sum_i \hat{n}_i \left( \hat{n}_i - 1 \right)+\sum_i \epsilon_i \hat{n}_i
\end{equation}

This provides an experimental framework within which to simulate the electron gas in solids, with the freedom to tune interparticle interactions by manipulating lattice parameters. Investigations may be performed into the effects of disorder, and within the Mott Insulator regime whereby atoms are localised onto individual lattice sites into topological states in the fractional quantum Hall effect, or even as a starting point for a quantum register~\cite{ref:BillyAndersonLoc, greiner_quantum_2002, ref:JakschFQHE, ref:ScalableComp}. However, the validity of the cold atom system as a quantum simulator strongly depends on the form of the trapping potential. The last term in Eq.~(\ref{eqn:BHM}) denotes an energy offset on individual lattice sites arising from a typically harmonic external confinement. The spatial dependence of the system density arising from the external potential~\cite{jaksch_cold_1998} associated with alternating superfluid and Mott-insulating regions is colloquially referred to as the `wedding-cake' structure. The resultant blurring of phase transitions makes comparison with theory less direct; eliminating this effect would allow answers to open questions such as the phase diagram of the fermionic Hubbard model~\cite{campo_quantitative_2007}. Precise control of the intensity distribution and elimination of external effects is of benefit to many other experimental situations including the observation of wave dynamics and quantum chaos using Bose-Einstein condensation confined in an optical corral-type potential~\cite{ref:corral}.

Compensation for the harmonic trapping term can be achieved either by direct cancellation with a compensatory potential to produce a uniform potential landscape, or by modifying the form of the trapping potential of experimental interest. Flat-topped beams hold particular appeal in this regard, used either as a square-well potential or to directly form standing wave optical lattices without a Gaussian envelope profile. Whatever the precise form of the chosen potential, an accurate and smoothly-varying beam profile is imperative to confining, manipulating and probing the trapped atoms.

	The freedom to meet the constraints imposed on the optical potential by experimental applications is granted by relaxing restrictions on the trapping plane phase. Optical trapping of ultracold atoms is facilitated by the dipole force, associated with a potential $U_{dip}\left(\vec{r}\right) \propto I(\vec{r})/\delta $ with $I(\vec{r})$ the spatially dependent laser intensity and $\delta$ the detuning of laser light from resonance. The dipole force therefore depends on the intensity gradient of the laser light. Any phase gradient affects only the scattering force, negligible under detuning far from the atomic resonance due to its $I/\delta^{2}$ dependence in comparison to the $I/\delta$ dependence of the dipole force.

Diffractive optical elements can be used to provide the necessary beam shaping precision and versatility. Using these techniques, we can obtain both continuous arbitrary intensity distributions and exotic lattice configurations inaccessible with standing wave interference alone such as circular distributions corresponding to an infinite 1D lattice~\cite{ref:ferriswheel, ref:ringlattices}. Our approach centres upon diffracting an incident laser beam using an acousto-optic deflector (AOD). The diffractive acoustic wave established in the AOD crystal is determined by a multiplexed input acoustic frequency signal generated using an arbitrary waveform synthesiser (AWS). The relative amplitudes within the multiplexed input determine the proportion of the total light diverted into the first diffracted order corresponding to the appropriate frequency, such that the total intensity pattern corresponds to the sum of the constituent diffracted beams. We thereby achieve precise control over both the position and amplitude of each diffracted beam. The application to discrete lattice patterns is evident, but by calculating the effect of neighbouring beam sites on each other, this approach can be easily extended to produce arbitrary composite continuous patterns including flat-topped beams. Alternatively, rather than the superposition of static frequencies, rapid deflection of a single beam such that trapped atoms experience a time-averaged potential has been successfully demonstrated in red-detuned potentials with minimal heating of trapped atoms~\cite{henderson_experimental_2009}. Dark optical lattices have also been realised by scanning around lattice sites~\cite{Arnold_BlueDetunedScanning}.  A combination of time-averaging and the composite beam approach presented here can yield a smoothly dynamically-varying potential, additionally enhancing the scalability of both methods. AOD-induced rotation and expansion of an optical lattice loaded with a BEC has been previously demonstrated~\cite{Accordion_OptExpress, AccordionAtoms}; dynamic shaped composite potentials would be a straightforward extension to this.
	
	A popular alternative approach to generating arbitrary and dynamic potentials is to use a computer-generated hologram imposed on an incident laser beam using a spatial light modulator (SLM), an array of either liquid crystal or micro-mirror pixels programmatically altering the beam phase. Improved versatility in the range of accessible potentials is granted the higher the phase resolution of each pixel, though this increases the complexity of the required numerical phase profile calculation. Iterative Fourier transform algorithms (IFTAs), of which multiple variants exist~\cite{ref:pasienski}, are extensively used to perform these high-resolution phase calculations. Pixel switching frequency is an important consideration if dynamic manipulation of trapped atoms is an experimental goal. The Texas Instruments digital micro-mirror device SLM has a switching frequency on the order of 50 \si{\kilo \hertz}, and a Boulder Nonlinear Systems ferroelectric liquid crystal SLM around 1 \si{\kilo \hertz}. However, these are both binary devices; with a phase resolution of $2\pi/256$, a Boulder Nonlinear Systems nematic liquid crystal SLM is capable of a far more versatile range of truly arbitrary potentials, but the switching frequency of hundreds of Hz could limit their applicability to dynamic manipulation of optical trapping potentials. AODs have an update frequency on the order of 10 \si{\mega \hertz}, facilitating almost seamless switching between dynamic frames, thus combining the versatility and switching rate necessary for an arbitrary dynamic manipulation sequence.

	In experimental situations without restrictions on the image plane phase, a significant advantage of using an AOD to form large continuous patterns is that the resultant potential is composed of multiple beams of different frequencies, the precise frequency separation dependent on the desired beam location and the details of the optical system. This frequency difference circumvents interference effects that arise when sculpting a single, highly coherent beam. Furthermore, unwanted beams arising from diffraction into the zeroth and higher orders are easily eliminated from the trapping plane intensity distribution, albeit with some loss of overall power. In contrast, achieving the highest-accuracy reproduction of large continuous arbitrary targets using spatial light modulation requires introduction of limited amplitude freedom in the trapping plane~\cite{ref:pasienski}, resulting in a significant noise accumulation near the trapping potential which can detrimentally perturb the experimental system.

We illustrate below the accuracy and versatility of the composite beam method for a range of continuous trapping potentials, and demonstrate a process by which an external potential can be compensated to produce a trapping potential tailored to specific experimental requirements. The first example compensates a harmonic term in the case of a square-well target potential; the applicability of the AOD beam shaping method to arbitrary continuous potentials is then illustrated, indicating the utility of the method in both creating and compensating arbitrary continuous potentials as required by the experimental conditions. Details of the experimental methods follow these examples.

\section{Beam shaping using an acousto-optic deflector}
	\subsection{Compensation potential}
	\label{section:comp}
	
	The effect of the additional harmonic confinement term is perhaps most immediately obvious with regard to flat-topped target potentials, although the principle of applying a compensation potential is identical in other cases. As discussed above, such flat-topped beams are experimentally applicable both in their own right and as a starting point for building other arbitrary potentials.
	
	To create a flat-topped beam using the superposition of diffracted beams, we initially consider the Sparrow resolution criterion~\cite{sparrow, ResolutionSurvey}. This refinement of the Rayleigh criterion is popular in astronomy, stating that multiple beams are indistinguishable, i.e. their composite intensity distribution perfectly flat, if the second derivative of this distribution is zero. The spacing $a$ between adjacent beams to achieve a flat-topped composite potential is therefore chosen such that:
	
\begin{equation}
	\label{eqn:sparrow}
	\frac{d}{dx} \big\{f(x)+f(x+a)\big\}=0 \quad \mbox{and} \quad \frac{d^2}{dx^2} \big\{f(x)+f(x+a)\big\}=0
\end{equation}
	In one dimension, our intensity distribution $f(x)$ is the sum of the $N$ constituent Gaussian beams, with $1/e^{2}$ waist $w$, and relative amplitudes  $A_{n}$ and positions $x_{n}$:

\begin{equation}
	\label{eqn:GaussSum}
	f(x) = \sum\limits_{n}^{N} A_{n} e^{-2(x-x_{n})^{2}/w^{2}}
\end{equation}
	
	The beam spacings calculated using the Sparrow criterion provide a good starting point for a feedback process that iteratively optimises beam spacings and relative amplitudes based on intensity variations measured across the composite beam profile; optimisation changes the frequency separations between beams by less than 10\% from their starting values in the case considered here.
	
\begin{figure}[ht]
	\centering
	\includegraphics[width=.8\textwidth]{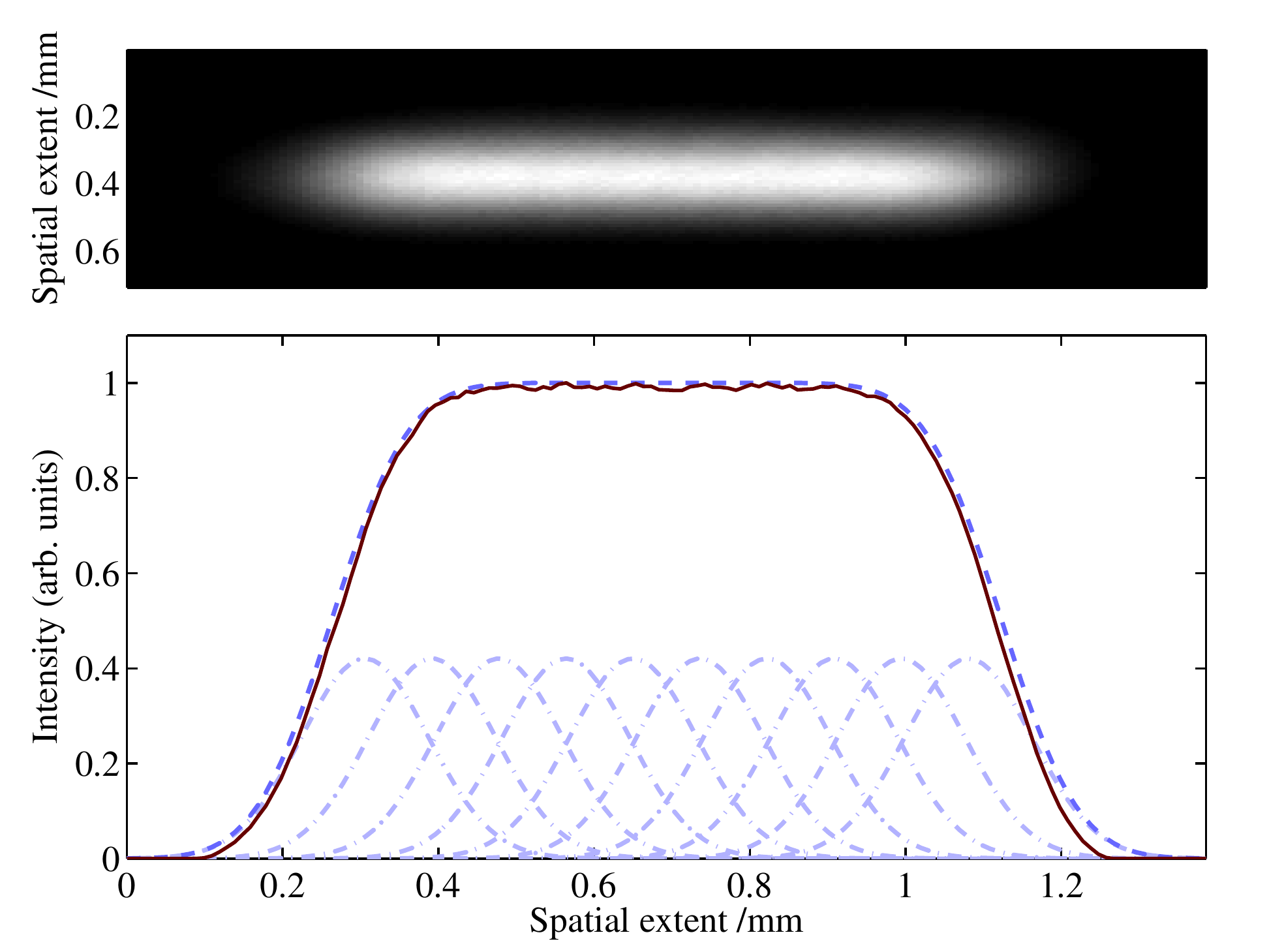}
	\caption{Intensity distribution measured using a CCD camera (top) for a flat-topped beam comprising 10 individual Gaussians of identical amplitude, with a line profile (bottom, solid line). Superposition of the dash-dot Gaussians yields the calculated target intensity distribution (dashed line).}
	\label{fig:flattop}
\end{figure}

	Figure~\ref{fig:flattop} shows the experimental realisation of a flat-topped intensity profile as the sum of 10 deflected beam components. In this case, the Sparrow criterion suggests a separation between adjacent beams of $a = 0.527 w$, with $w$ the beam waist. After optimisation, the experimental error is 1.4\% over the flat region of the intensity profile and 2.3\% over the full distribution. In this and subsequent figures, the corrugations visible on the compensation potential arise from dust specks on the imaging camera rather than being features of the potential itself.

	Using a sequence of equal-amplitude constituent beams, the composite potential has a power-law dependence with a maximum order scaling approximately with the number of beams used. The general form of the intensity profile is:

\begin{equation}
	\label{eqn:powerlaw}
	I(x) = \sum_{n} A_{n} x^{n}
\end{equation}

The beam shown in Fig.~\ref{fig:flattop} is associated with a potential proportional to $x^{10}$. Whilst the example shown is for a single row of Gaussians, this principle can be readily extended to constructing arbitrary-shaped two-dimensional potentials by deflecting the beam in both x- and y-directions with separations calculated as above (see the Experimental Techniques section for discussion of the dual-axis AOD). The only subtlety arising from additional rows of beams is that if the frequency spacing is equal in both x- and y-directions, then beams lying along diagonal lines have the same frequency and thus interfere. Such undesirable interference is easily avoided by using different frequency spacings and elliptical spots to fulfill the Sparrow criterion in both directions; elliptical distributions occur anyway in the focal plane of an optical system with a high numerical aperture and linearly polarised light~\cite{ref:ellipticalspot}. This extension is simple in comparison to a similar extension of the target output of a computer-generated hologram. With an IFTA used to improve the range and versatility of accessible patterns, hologram calculation becomes more complicated for large continuous potentials due to the appearance of optical vortices in the calculation process~\cite{ref:senthilkumaran}. Furthermore, limited by a finite pixel array, spatial light modulators find it difficult to realise a sharp edge to a flat-topped beam due to the high Fourier-space frequencies required. A super-Lorentzian target array is therefore often used, the order of which is a compromise between flatness and calculation accuracy. In contrast, the AOD composite beam approach has an intensity falloff limited only by the beam waist. For example, using a composite flat-top consisting of 10 beams as in Fig.~\ref{fig:flattop}, the intensity falls from 95\% of its maximum value to 5\% over $1.6 w$, whereas an eighth-order super-Lorentzian as demonstrated in~\cite{liang_high-precision_2010}, of identical width, has the same intensity falloff over $2.0 w$. The accuracy of exotic patterns calculated using an IFTA can be improved by incorporating Helmholtz propagation into hologram calculation~\cite{ref:gaunt}; although this does not as yet match the approximately 1\% RMS error setting the current accuracy limit on a flat-top generated by binary spatial light modulation~\cite{liang_high-precision_2010}, binary devices are associated with intensity profiles of restricted complexity. This SLM accuracy limit is slightly higher than the accuracy obtained using the AOD composite beams above, but this small difference should be balanced against the improved edge definition and greater complexity possible with the AOD as opposed to a binary SLM.

However, these flat-topped beams are unlikely to be used in isolation. A harmonic term arising from the external potential of a magnetic trap or additional dipole trapping beam will typically dominate over higher-order power-law terms unless compensated. The flexibility of the composite beam AOD approach allows such compensation to be implemented straightforwardly, with a potential that cancels out the dominant low-order terms of the power-law Taylor expansion.

An arbitrary potential along the x-axis can be expressed as a Taylor series up to $n_{max}$, the order we want to cancel:
 \begin{equation}
 	\label{eqn:extTaylor}
	V_{tot}(x)=\sum_{n=0}^{n_{max}} \frac{V_{tot}^{(n)}(x_0)}{n!}(x-x_0)^n+\mathcal{O}(n_{max}+1)
\end{equation}
We fit the functional form of this potential using a sum of equal-width Gaussians by adjusting their relative positions and amplitudes. The accuracy increases with the number of beams used, and depends on the complexity of the target distribution. Along the same axis this AOD-generated composite potential has the form:
 \begin{equation}
V_{dip}(x)=\sum_{n=1}^N a_i V_{beam}(x-s_i)
\end{equation}
where $N$ is the number of constituent beams, $a_i$ the amplitude of each beam, $s_i$ the displacement along the x-axis and
$V_{beam}(x)$ the Gaussian function produced by each constituent deflected beam. The optimal set of parameters ${a_{i}, s_{i}}$ to cancel the external potential are determined using an optimisation routine.

\begin{figure}[ht]
	\centering
	\includegraphics[width=.8\textwidth]{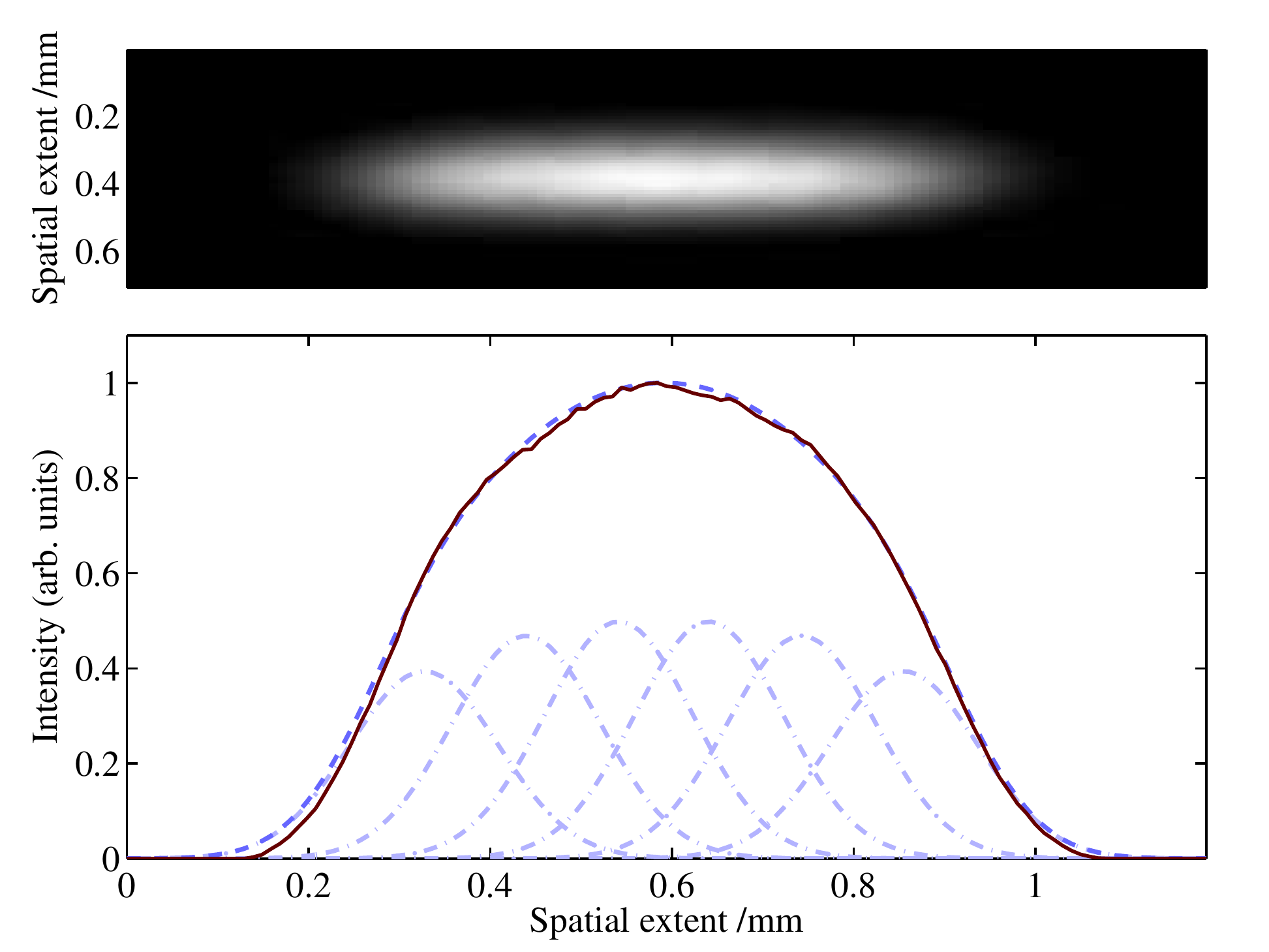}
	\caption{Intensity distribution measured using a CCD camera (top) and the corresponding line profile (bottom, solid line) for a harmonic compensation potential, with the target intensity distribution (dashed line) the sum of the dash-dot Gaussians.}
	\label{fig:compensation}
\end{figure}

	Figure~\ref{fig:compensation} illustrates the use of 6 beams to cancel an $\mathcal{O}\left(2\right)$ term, with matching performed using the Taylor expansion of the potentials. These beams would have a blue frequency detuning to create a repulsive potential. For this example, the experimental error over the entire pattern is 1.8\%. 

	Smoothing the nonuniform density distribution of an ultracold gas cloud trapped in an optical box has been experimentally demonstrated using repulsive spots of laser light positioned using an acousto-optic modulator at points along the axis of the cloud~\cite{meyrath_bose-einstein_2005}. This illustrates the viability of the compensation method in correcting small-scale beam imperfections that can fragment the cloud. However, the current approach and potential generated in Fig.~\ref{fig:compensation} focusses primarily on offering compensation for large-scale external or residual continuous harmonic potentials perturbing the overall form of the trapping potential. As in~\cite{meyrath_bose-einstein_2005}, these may be superimposed onto a dipole potential to smooth out the residual confinement terms and provide a uniform potential landscape, but the combination of this compensation with continuous beam shaping also allows the trapping laser beams to be directly modified.
	
	\subsection{Arbitrary continuous potentials}
	\label{section:buchleitner}
	
\begin{figure}[ht]
	\centering
	\includegraphics[width=.8\textwidth]{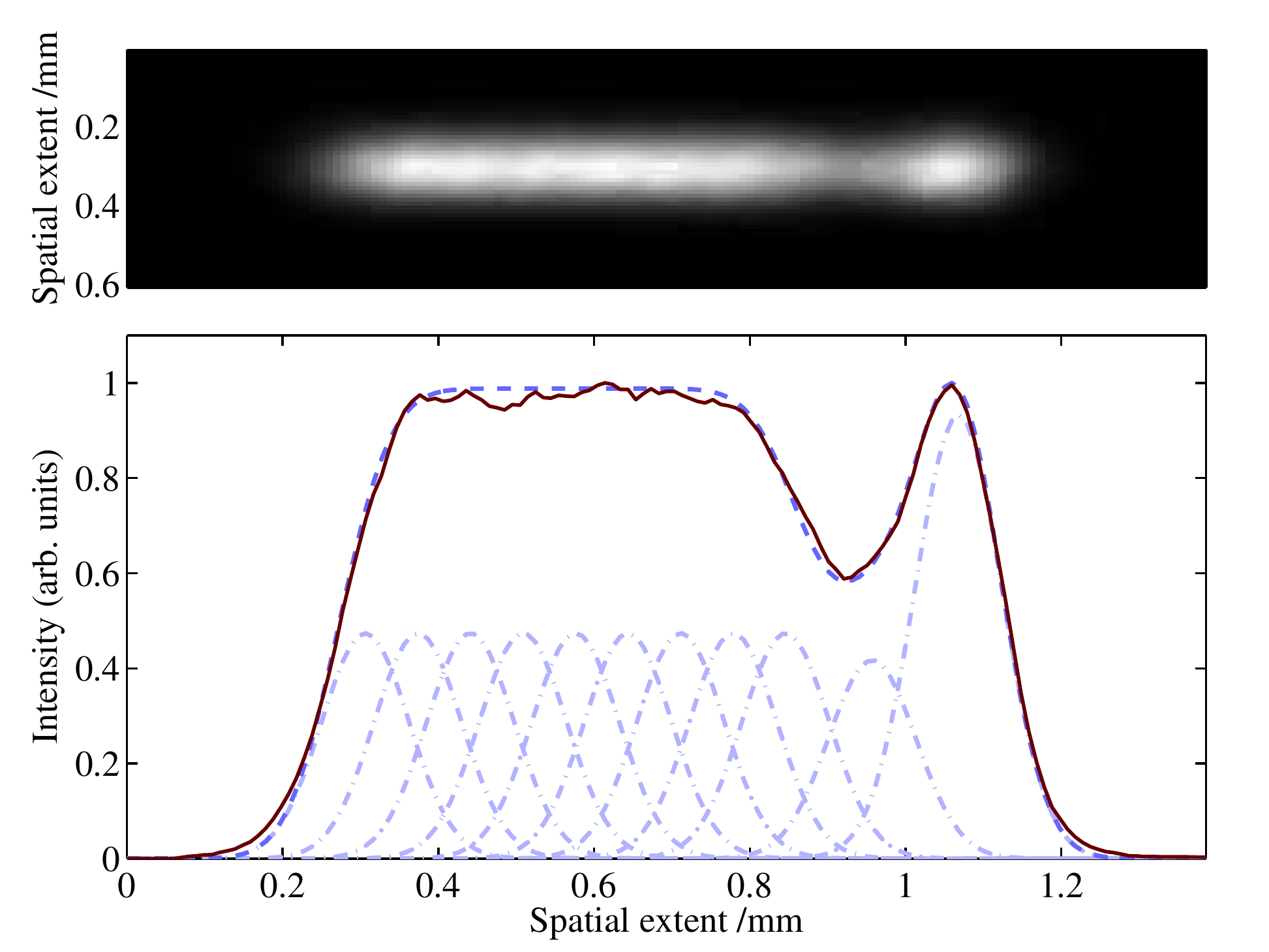}
	\caption{Intensity distribution measured using a CCD camera (top) and corresponding line profile (bottom, solid line) for an extension of the flat-topped beam creating a broad reservoir connected to a single well, with applications to studying decoherence in quantum systems. The target intensity (dashed line) is the sum of the dash-dot Gaussians.}
	\label{fig:buchleitner}
\end{figure}

	As indicated by the ability to  modify the target to incorporate a compensation term, this beam sculpting approach can be applied to generating arbitrary discrete or continuous trapping potentials. The example illustrated in Fig.~\ref{fig:buchleitner} extends the flat-topped beam of Fig.~\ref{fig:flattop} by the addition of a spatially separated single Gaussian, to form a potential analogous to a single well connected to a reservoir of variable size. This type of potential has been used for theoretical studies of tunneling and decoherence~\cite{ref:Buchleitner} and our method is well-suited to realising this in practice.

The flat-topped reservoir of Fig.~\ref{fig:buchleitner} consists of 9 deflected beams, optimised from the starting point of equal amplitude and a Sparrow separation of $0.552 w$. The well depth is controlled by a low-amplitude Gaussian midway between the reservoir and a spatially separated single well. The parameters defining the potential and the interplay between reservoir and single well are sufficiently flexible that the intensity distribution can be easily and precisely modified, allowing dynamic real-time manipulation of trapped atoms. The illustrated experimental realisation has an error of 2.1\% over the entire pattern region.

This example illustrates the utility of this method in generating both arbitrary, non-symmetric continuous potentials, and single diffraction-limited points that could be arranged in a discrete lattice structure. Although this method is most suitable for shapes that can be expressed as sums of Gaussians, this is not a significant limitation: numerous arbitrary intensity distributions can be generated with a high level of accuracy. The reproduction accuracy of all patterns could be significantly improved with further optimisation.

\section{Conclusion}

The power of the composite beam method lies in its simplicity. With rudimentary optimisation based on the measured intensity profile, large continuous potentials, or indeed discrete patterns, may be reproduced without loss of accuracy resulting from interference or increased numerical calculation complexity. Arbitrary patterns, most notably flat-topped beams, can be produced with an error of around 2\%, comparing favourably to results demonstrated using SLM techniques, and without compromising either versatility or frame update rate. Systematic optimisation would further enhance this accuracy. Although approached from the opposite direction, superposition of constant frequencies is comparable to the rapid painting of a single beam presented in~\cite{henderson_experimental_2009} in terms of the access to arbitrary continuous potentials with high reproduction accuracy; furthermore, the scalability of each of these methods could be enhanced by their combination.

Alongside other examples, the versatility of the approach has been utilised in producing a repulsive compensation potential, which could be used either to directly shape a target potential, or to create a flat background on which to construct additional potentials. This could significantly improve the relation of quantum simulation experiments to theoretical calculations, allowing phase transitions to be investigated with greater clarity. The compensation process could also be applied in SLM-based experiments, although measures would have to be employed to restrict vortex formation in large continuous distributions. Whilst this investigation focussed upon flat-topped and compensatory potentials, and a uniform continuous potential for decoherence investigations, the precise shaping inherent in the composite beam approach holds universal appeal in creating arbitrary discrete and continuous trapping potentials for a wide range of processes.

\appendix
\section{Experimental techniques}
	\label{sec:methods}

	Beam shaping is performed using a dual-axis AOD (Isomet LSA110A-830XY). Aligned to a central frequency of 45 MHz, angular deviations in either the x- or y-directions are determined according to 
\begin{equation}
	\label{eqn:angle}
	\Delta \theta = \frac{\Delta f \lambda}{u_{s}}
\end{equation}
where $u_{s} \approx 610$ ms$^{-1}$ is the speed of sound in the TiO$_{2}$
 crystal of the AOD, $\Delta f$ the frequency deviation and $\lambda$ the wavelength of light used: 780 nm in section~\ref{section:comp} and 830 nm in section~\ref{section:buchleitner}. We determined $u_{s}$ from the measured angular separation of the diffracted beams for given input frequencies. The resultant linear separation of spots corresponding to different diffraction angles is fixed by a lens placed at a focal length from the AOD, serving to focus the spatially separated beam components onto a CCD camera (Unibrain Fire-i 521b) as well as ensuring parallel propagation to this point. For imaging purposes, this focussing lens was of focal length $f = 400$ mm in section~\ref{section:comp} and $f = 500$ mm in section~\ref{section:buchleitner}; with lens diameters of 25.4 mm, these yield diffraction-limited beam sizes of 162 \si{\micro\metre} and 203 \si{\micro\metre} respectively. To apply this technique to dipole force traps requires significantly smaller diffraction-limited beam sizes; for example, an objective lens with a focal length of 40 mm, diameter 25.4 mm and corresponding diffraction-limited beam size of 1.6 \si{\micro\metre} with wavelength 830 nm produces a trapping frequency of a few kHz for $^{87}$Rb atoms. This is well below the hundreds of kHz acoustic deflection frequency difference between constituent beams of the composite potential such that any beating between neighbouring spots will not adversely affect atoms with the trapping frequencies considered here. 
 
 	Relative amplitudes $A_{n}$ and image-plane positions $(x_{n},y_{n})$ of the multiple spots comprising the sculpted beam are computationally determined by fitting a function corresponding to Eq.~\ref{eqn:GaussSum}, a sum of Guassian beams, to a target intensity array with the Sparrow criterion providing a useful starting point for flat-topped beams. A small amount of manual optimisation is performed on the experimental output, although the results are likely to benefit from an automated optimisation routine as has been demonstrated in conjunction with an IFTA for holographic beam shaping~\cite{ref:SLMfeedback}. The frequencies corresponding to the necessary relative separations are calculated, and then imposed upon the AOD using an arbitrary waveform synthesiser (AWS) (Hewlett Packard 8770A). Granting independent control of the relative amplitudes $a_{i}$ of its constituent frequency components $f_{i}$, the AWS output resembles a multiplexed signal of the form:
\begin{equation}
	\label{eqn:AWS}
	f_{out} = \sum\limits_{i = 1}^{N}a_{i} f_{i}
\end{equation}
	
	The AWS signal (with a maximum output of 1 dBm) is amplified using a 1.6 W amplifier (Motorola CA2832C). 
	The calculated frequencies are transmitted to the AWS internal memory using a Labview GPIB interface; the frequency is incorporated in the expression for the output signal via the input value $n$ according to the internal clock frequency $f_{clock} = 125$ MHz:
	
\begin{equation}
	\label{eqn:AWSf}
	f = \left( \frac{n}{\text{number of elements}} \right) \times  f_{clock}
\end{equation}

A series of spatially separated spots requires an output signal consisting of all appropriate frequencies at the correct relative amplitudes to create the multiplexed standing wave diffraction pattern in the AOD. The GPIB signal takes the form:

\begin{equation}
	\label{eqn:AWSsignal}
	S_{in} = \sum_{i} a_{i} \sin\left(\frac{2 \pi n_{i}}{L}\right)
\end{equation}
where $a_{i}$ is the amplitude associated with each frequency $f_{i}$, with the total amplitude less than 2048, and $L$ the total packet length. The packet length is the number of total number of points per period multiplied by the total number of periods in each wave segment, and must be less than 5320. The number of points per period is a compromise between being a multiple of 8 and a multiple of the time period associated with the lowest beating frequency in the signal as a fraction of the internal clock period. This latter condition minimises flicker at the start of each wave segment loop. The number of deflecting frequencies output through the AWS at a single time can limit the complexity of the composite intensity profile. However, the examples illustrated above demonstrate that even a restricted number of deflected beams can be used to accurately reproduce a range of potentials.

This work was supported by EPSRC (Grant No. EP/J008028/1) and DT was supported by the Bodossaki Foundation and St. Peter's College, Oxford.

\end{document}